\documentclass[11pt]{article}
\usepackage[dvips]{epsfig}
\usepackage{epsfig,psfig,latexsym,citesort}

\input{amssym.def}
\input{amssym}

\newtheorem{lemma}{Lemma}
\newcommand{\qed}{~$\vrule width.15cm height.2cm depth0cm$ \medbreak}
\newenvironment{proof}{\noindent{\bf Proof: }}{\qed}

\newcommand {\mm}[1]{\ifmmode{#1}\else{\mbox{\(#1\)}}\fi}

\newcommand{\real}{\mm{{\Bbb R}}}

\newcommand{\utwi}[1]{\mbox{\boldmath $ #1$}}
\newcommand{\ba}{{\utwi{a}}}

\newcommand{\bc}{{\utwi{c}}}

\newcommand{\bs}{{\utwi{s}}}

\newcommand{\bw}{{\utwi{w}}}
\newcommand{\bx}{{\utwi{x}}}
\newcommand{\by}{{\utwi{y}}}

\begin{document}
     \title{\bf On Design of Optimal Nonlinear Kernel Potential Function for
Protein Folding and Protein Design }

      \author{\bf Changyu Hu, Xiang Li and Jie Liang\thanks{Corresponding author.  Phone:
      (312)355--1789, fax: (312)996--5921, email: {\tt
      jliang@uic.edu}} \\ Department of Bioengineering, SEO, MC-063 \\
      University of Illinois at Chicago\\ 851 S.\ Morgan Street, Room
      218 \\ Chicago, IL 60607--7052, U.S.A.}  \date{\today}

      \maketitle

\abstract{ 
Potential functions are critical for computational studies of protein
structure prediction, folding, and sequence design.  A class of widely
used potentials for coarse grained models of proteins are contact
potentials in the form of weighted linear sum of pairwise contacts.
However, these potentials have been shown to be unsuitable choices
because they cannot stabilize native proteins against a large number
of decoys generated by gapless threading, when the number of native
proteins is above 300.  We develop an alternative framework for
designing protein potential. We describe how finding optimal protein
potential can be understood from two geometric viewpoints, and we
derive nonlinear potentials using mixture of Gaussian kernel functions
for folding and design.  In our experiment we use a training set of
440 protein structures repre senting a major portion of all known
protein structures, and about 14 million structure decoys and sequence
decoys obtained by gapless threading.  The optimization criterion for
obtaining parameters of the potential is to minimize bounds on the
generalization error of discriminating protein structures and decoys
not used in training.  We succeeded in obtaining nonlinear potential
with perfect discrimination of the 440 native structures and native
sequences.  For the more challenging task of sequence design when
decoys are obtained by gapless threading, we show that there is no
linear potential with perfect discrimination of all 440 native
sequences.  Results on an independent test set of 194 proteins also
showed that nonlinear kernel potential performs well, with only 3
structures and 14 sequences misclassified, which compare favorable
with the results of 7 structures and 37 sequences misclassified using
optimal linear potential.  We conclude that more sophisticated
formulation other than the simple weighted sum of contact pairs can be
useful.
}
\vspace*{1in}


\newpage
\section{Introduction}

Potential function plays critical roles in computational studies of
protein folding, protein structure prediction, and protein sequence
design \cite{Dill95_PS,Levitt97_ARB,Bonneau01_ARBBS,Saven01_CR}.  
A variety of empirical potential functions have been developed for
coarse-grained models of proteins, where amino acid residues are not
represented at atomic details.  One prominent class of potentials are
knowledge-based potentials derived from statistical analysis of
database of protein structures
\cite{TanakarScheraga76,Miyazawa85_M,SamudralaMoult98_JMB,LuSkolnick01_Proteins}.
In this class of potentials, the interaction between a pair of
residues are estimated from its relative frequency in database when
compared with a reference state or a null model.  This approach has
been successfully applied in fold recognition, in threading, and in
many other studies
\cite{Miyazawa96_JMB,SamudralaMoult98_JMB,LuSkolnick01_Proteins,Wodak93_COSB,Sippl95_COSB,Lerner95_Proteins,Jernigan96_COSB,Simons99_Proteins}.
The drawback of this class of potential is that there are several
conceptual difficulties. These include the neglect of chain
connectivity in the reference state, and the problematic implicit
assumption of Boltzmann distribution
\cite{ThomasDill96_JMB,ThomasDill96_PNAS,Ben-Naim97}.  An alternative
approach is to find a set of parameters such that the potential
functions are optimized by some criterion, {\it e.g.}, maximized
energy difference between native conformation and a set of alternative
(or decoy) conformations
\cite{Goldstein92_PNAS,MaiorovCrippen92_JMB,ThomasDill96_PNAS,TobiElber00_Proteins_1,Vendruscolo98_JCP,Vendruscolo00_Proteins,Bastolla01_Proteins,Dima00_PS,Micheletti01_Proteins}.
This approach has been shown to be very effective in recognizing
native structures from alternative conformations
\cite{Micheletti01_Proteins} and in folding membrane proteins
\cite{Dobbs02_Proteins}.  However, if a large number of native protein
structures are to be simultaneously stabilized against a large number
of decoy conformations, no such potential functions can be found
\cite{Vendruscolo00_Proteins,TobiElber00_Proteins_1}.

There are three key steps in developing effective empirical potential
function using optimization: (1) the functional form of the potential,
(2) the generation of a large set of decoys for discrimination, and
(3) the optimization techniques.  The initial step of choosing an
appropriate functional form so far has been straightforward.
Empirical pairwise potentials are usually all in the form of weighted
linear sum of interacting residue pairs (see reference
\cite{Fain02_PS} for an exception).  In this functional form, the
weight coefficients are the parameters of the potential, which are
optimized for discrimination.  The same functional form is also used
in statistical potential, where the weight coefficients are derived
from database statistics.  For the task of decoy generation, an
efficient method to obtain millions of decoys is gapless threading,
{\it i.e.}, a decoy conformation can be obtained by mounting the
sequence of the native protein to different parts of the conformation
of a larger protein of an unrelated structure
\cite{MaiorovCrippen92_JMB}.  Alternatively, a large number of
challenging decoys can be generated by using either a chain growth
method \cite{Vendruscolo97_FD,Vendruscolo98_FD} or a protein structure
refinement method \cite{Monster_Skolnik_JMB_97}, both are technically
more complex.  The optimization techniques that have been used include
perceptron learning and linear programming
\cite{TobiElber00_Proteins_1,Vendruscolo00_Proteins}, and analytical
solution has also been obtained when assumptions about the
distribution of data are made \cite{XiaLevitt00_JCP}.  The objectives
of optimization are often maximization of energy gap between native
conformation and the average of decoy conformation, or energy gap
between native and decoys with lowest energy, or the $z$-score of the
native structure
\cite{Goldstein92_PNAS,Koretke96,Koretke98_PNAS,Hao96_PNAS,MirnyShkh96_JMB}.

In this study, we develop an alternative formulation of protein
potential function in the form of mixture of nonlinear Gaussian kernel
functions.  We also use a different optimization technique based on
quadratic programming.  Instead of maximizing the energy gap, here a
function related to bounds of expected classification errors is
optimized 
\cite{VapChe74,Vapnik95,Burges98,ScholkopfSmola02}.  
In addition, we explore
the relationship between protein folding and protein design.  We study
a simplified version of the protein folding problem.  Our goal is to
identify the native protein structure from an ensemble of alternative
or decoy structures for a given amino acid sequence.  We also study a
simplified version of the protein design problem.  Our goal of protein
sequence design is to identify a protein sequence that is most
compatible with a given three-dimensional coarse-grained structure.
In this study, we do not address the problem of how to generate
candidate conformation or candidate sequence by searching either the
conformation space or the sequence space.

Experimentation with the nonlinear function developed shows that it
can discriminate simultaneous 440 native proteins against 14 million
structure decoys generated by gapless threading.  Results of similar
experiments with perceptron learning was negative, as reported in
literature \cite{Vendruscolo00_Proteins}.  We also test our potential
function for protein design.  Using the same training set of 440
proteins and a corresponding set of 14 millions sequence decoys, we
succeed in developing a nonlinear function that correctly classifies
all 440 native sequences. In contrast, we cannot obtain a weighted
linear sum potential using the state-of-the-art interior point solver
of linear programming method as reported in
\cite{TobiElber00_Proteins_1,Meller02_JCC}, such that it is capable of
classifying perfectly all the native sequences.  We also perform blind
tests for both native structure and native sequence recognition.
Taking 194 proteins unrelated to the 440 training set proteins, the
nonlinear potential achieves a success rate of 92.8\% and 98.4\% in
sequence design and in structure recognition, respectively.  Both
results compare favorably with optimal linear potential and
statistical potential.

The rest of the paper is organized as follows.  We first describe
theory and model of linear and nonlinear function, including the kernel
model and the optimization technique.  We then explain details of
computation.  We further describe experimental results of learning
and results of blind test.  We conclude with discussion.

\section{Theory and Models}

\paragraph{Modeling Protein Folding Potential.}
To model protein computationally, we first need a method to describe
its geometric shape and its sequence of amino acid residues.
Frequently, a protein is represented by a $d$-dimensional vector $\bc
\in \real^d$.  For example, we can represent a protein as a vector
$\bc \in \real^d$, $d=20$, by measuring the solvent accessible surface
areas of each of the 20 residue types.  A method that is widely used
is to count nonbonded contacts of various types of amino acid residue
pairs in a protein structure.  In this case, the count vector $\bc \in
\real^d, d=210$, is used as the protein descriptor.  Once the
structural conformation of a protein $\bs$ and its amino acid sequence
$\ba$ is given, the protein description $f: (\bs, \ba) \mapsto
\real^d$ will fully determine the $d$-dimensional vector $\bc$.  In
the case of contact vector, $f$ corresponds to the mapping provided
by specific contact definition, {\it e.g.}, two residues are in
contact if their distance is below a specific cut-off threshold
distance.

Based on the classical experiments of Anfinsen
\cite{Anfinsen_Nature_1973}, a fundamental requirement for protein
folding is that the native structure $\bs_N$ with native amino acid
sequence $\ba_N$ must have an energy $H(f(\bs_N, \ba_N))$ that is the
lowest among a set of alternative structure called decoys ${\cal D} =
\{(\bs_D, \ba_N)\}$, where the native amino acid sequence $\ba_N$ takes a decoy
conformation $\bs_D$ that is different from the native conformation
$\bs_N$ \cite{MaiorovCrippen92_JMB,Jones92_Nature}:
\[
H(f(\bs_N, \ba_N)) < H(f(\bs_D, \ba_N)) \quad \mbox{for all } (\bs_D, \ba_N)  \in {\cal D}
\]
Sometimes we can further require that the energy difference must be
greater than a constant $b>0$:
\[
H(f(\bs_N, \ba_N)) + b < H(f(\bs_D, \ba_N)) \quad \mbox{for all }
(\bs_D, \ba_N) \in {\cal D}
\]

A widely used functional form for protein potential function $H$ is
the weighted linear sum of pairwise contacts
\cite{TanakarScheraga76,Miyazawa85_M,TobiElber00_Proteins_1,Vendruscolo98_JCP,SamudralaMoult98_JMB,LuSkolnick01_Proteins}.
The linear sum energy $H$ is then:
\begin{equation}
H(f(\bs, \ba)) = H(\bc) = \bw \cdot \bc.
\label{linear}
\end{equation}
As soon as the weight vector $\bw$ is specified, the potential
function is fully defined.  For such linear potentials, the basic
requirement for protein potential is then:
\[
\bw \cdot (\bc_N - \bc_D) < 0
\]
or 
\[
\bw \cdot (\bc_N - \bc_D) + b < 0
\]
if we require that the energy difference between a native
structure and a decoy must be greater than a real value $b$.

\paragraph{Two Geometric Views of Linear Protein Folding Potentials.}

\begin{figure}[tbh]
     \centerline{\epsfig{figure=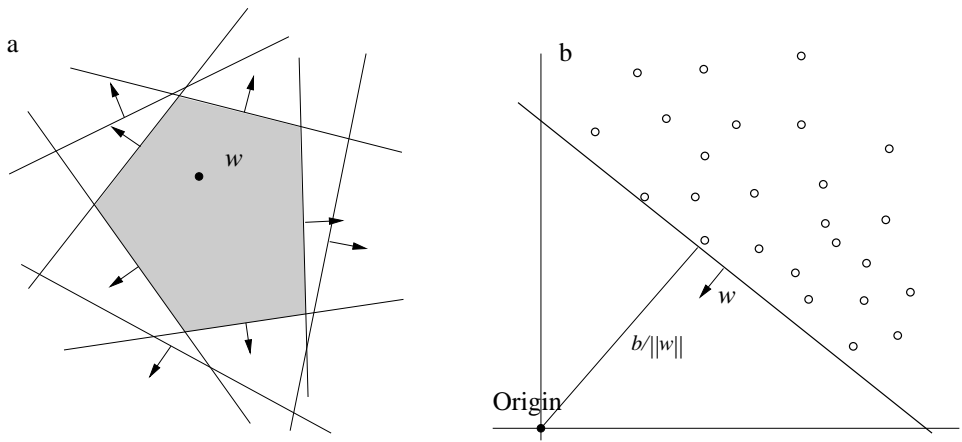,width=5.in}}
\caption{\sf
Geometric views of the inequality
requirement for weighted linear protein potential. (a). In the first
geometric view, the space $\real^d$ is divided into two half-spaces by
the hyperplane $\bw \cdot (\bc_N - \bc_D) + b < 0$. The hyperplane is
defined by the normal vector $(\bc_N - \bc_D)$ and its distance
$b/||\bc_N - \bc_D ||$ from the origin.  The weight vector must be
located in the half-space opposite to the direction of the normal
vector $(\bc_N - \bc_D)$.  (b). A second geometric view of the
inequality requirement for linear protein potential.  The space
$\real^d$ is divided into two half-spaces by the hyperplane $\bw \cdot
(\bc_N - \bc_D) + b < 0$. Here the hyperplane is defined by the normal
vector $\bw$ and its distance $b/||\bw ||$ from the origin.  All
points $\{\bc_N - \bc_D\}$ are located on one side of the hyperplane
away from the origin.
}
\vspace*{.1in}
\label{Fig:geom1}
\end{figure}

There is a natural geometric view of the inequality requirement for weighted
linear sum potentials.  A useful observation is that each of the
inequalities divides the space of $\real^d$ into two halfs separated
by a hyperplane. The hyperplane is defined by the normal vector
$(\bc_N - \bc_D)$ and its distance $b/||\bc_N - \bc_D ||$ from the
origin.  The weight vector $\bw$ must be located in the half-space
opposite to the direction of the normal vector $(\bc_N - \bc_D)$
(Fig~\ref{Fig:geom1}a).  This half-space can be written as $\bw \cdot
(\bc_N - \bc_D) + b < 0$.

When there are many inequalities to be satisfied simultaneously, the
intersection of the half-spaces forms a convex polyhedron
\cite{Edels87}. If the weight vector is located in the polyhedron, all
the inequalities are satisfied.  Potentials with such weight vector
$\bw$ can discriminate the native protein from the set of all decoys.

For each native protein $i$, there is one convex polyhedron ${\cal
P}_i$ formed by the set of inequalities associated with its decoys.
If our potential can discriminate simultaneously $n$ structures from a
union of sets of decoys, the weight vector $\bw$ must be located in a
smaller convex polyhedron $\cal P$ that is the intersection of the $n$
convex polyhedra:
\[
\bw \in {\cal P} = \bigcap_{i=1}^n
{\cal P}_i.
\]

There is yet another geometric view of the inequality requirements.
The relationship $\bw \cdot (\bc_N - \bc_D) + b < 0$ for all decoys
and native protein structures can be regarded as a requirement that all
points $\{\bc_N - \bc_D\}$ are located on one side of a hyperplane,
which is defined by its normal vector $\bw$ and its distance
$b/||\bw||$ to the origin (Fig~\ref{Fig:geom1}b).  We can show that
such a hyperplane exists if the origin is not contained within the
convex hull of the set of points $\{ \bc_N - \bc_D\}$
(see Appendix).

The second geometric view is dual and mathematically equivalent to the
first geometric view.  In the first view, a point $\bc_N - \bc_D$
determined by the structure-decoy pair $c_N = (\bs_N,\ba_N)$ and
$c_D=(\bs_D, \ba_N)$ corresponds to a hyperplane representing an
inequality, a solution weight vector $\bw$ corresponds to a point
located in the final convex polyhedron.  In the second view, each
structure-decoy pair is represented as a point $ \bc_N -\bc_D$ in
$\real^d$, and the solution weight vector $\bw$ is represented by a
hyperplane separating all the points ${\cal C} = \{\bc_N - \bc_D \}$
from the origin.

\paragraph{Optimal Linear Potentials.}
Several optimization methods have been applied to find the weight
vector $\bw$.  The Rosenblantt perceptron method works by iteratively
updating an initial weight vector $\bw_0$
\cite{Vendruscolo98_JCP,Micheletti01_Proteins}.  Starting with a random vector, {\it
e.g.}, $\bw_0 ={\bf 0}$, we test each native protein and its decoy
structure.  Whenever the relationship $ \bw \cdot (\bc_N - \bc_D) + b
< 0$ is violated, we update $\bw$ by adding to it a scaled vector
$\eta \cdot (\bc_N - \bc_D)$.  The final weight vector is therefore a
linear combination of protein and decoy count vectors:
\begin{equation}
\bw = \sum \eta \cdot (\bc_N - \bc_D) = \sum_{N \in {\cal N}} \alpha_N
\cdot \bc_N - \sum_{D \in {\cal D}} \alpha_D \cdot \bc_D.
\label{LinearEq}
\end{equation}
Here $\cal N$ is the set of native proteins, and $\cal D$ is the set
of decoys.  The set of coefficients $\{\alpha_N \} \cup \{\alpha_D\}$
gives a dual form representation of the weight vector $\bw$ as an
expansion of the training examples, including both native and decoy
structures.

If the final convex polyhedron $\cal P$ is non-empty, there can be
infinite number of choices of $\bw$, all with perfect discrimination.
But how do we find a weight vector $\bw$ that is optimal?  This
depends on the criterion for optimality.  For example, one can choose
the weight vector $\bw$ that minimizes the variance of energy gaps
between decoys and natives: $ \arg_\bw \min \frac{1}{|{\cal D}|} \sum
\left( \bw \cdot (c_N - c_D)\right)^2 - \left[ \frac{1}{|{\cal
D}|}\sum_D \left(\bw \cdot (\bc_N -\bc_D)\right) \right]^2 $ as used in
reference \cite{TobiElber00_Proteins_1}, or minimizing the $Z$-score
of a large set of native proteins, or minimizing the $Z$-score of the
native protein and an ensemble of decoys
\cite{ChiuGoldstein98_FD,MirnyShkh96_JMB}, or maximizing the ratio $R$
between the width of the distribution of the energy and the average
energy difference between the native state and the unfolded ones
\cite{Goldstein92_PNAS,Hao99}.
A series of important works using perceptron learning and other
optimization techniques
\cite{FriedrichsWolynes89_Science,Goldstein92_PNAS,TobiElber00_Proteins_1,Vendruscolo98_JCP,Dima00_PS}
showed that effective linear sum potentials can be obtained.

Here we describe yet another optimality criterion according to the
second geometric view.  We can choose the hyperplane $(\bw, b)$ that
separates the points $\{\bc_N - \bc_D\}$ with the largest distance to
the origin.  Intuitively, we want to characterize proteins with a
region defined by the training set points $\{\bc_N - \bc_D \}$.  It is
desirable to define this region such that a new unseen point drawn
from the same protein distribution as $\{\bc_N - \bc_D \}$ will have a
high probability to fall within the defined region, and non-protein
points following a different distribution, which is assumed to be
centered around the origin when no {\it a priori\/} information is
available, will have a high probability to fall outside the defined
region.  In this case, we are more interested in modeling the region
or support of the distribution of protein data, rather than estimating
its density distribution function.  For linear potential, regions are
half-spaces defined by hyperplanes, and the optimal hyperplane $(\bw,
b)$ is the one with maximal distance to the origin.  This is related
to the novelty detection problem and single-class support vector
machine studied in statistical learning theory
\cite{VapChe64,VapChe74,ScholkopfSmola02}.  In our case, any
non-protein points will need to be detected as outliers from the
protein distribution characterized by $\{\bc_N - \bc_D \}$.  Among all
linear functions derived from the same set of native proteins and
decoys, the optimal weight vector $\bw$ is likely to have the least
amount of mislabellings.
The optimal weight vector $\bw$ can therefore be found by solving the
following primal quadratic programming problem:
\begin{eqnarray*}
\mbox{Minimize } & \frac{1}{2} || \bw||^2
\\
\mbox{subject to} & \bw \cdot (\bc_N - \bc_D) + b < 0 \mbox{ for all }
N \in {\cal N} \mbox{ and } D \in {\cal D}.
\end{eqnarray*}
The solution maximizes the distance $b/||\bw||$ of the plane $(\bw,
b)$ to the origin.  
The dual form of the same quadratic programming
problem can be written as \cite{Mangasarian94}:
\begin{eqnarray*}
\mbox{Minimize } & \sum_{i,j} \alpha_i \alpha_j \cdot 
\bx_i \cdot \bx_j
\\
\mbox{subject to} & \alpha_i \ge 0 \mbox{ and } \sum_i \alpha_i = 1
\end{eqnarray*}
where  $\bx_i, \bx_j \in  \{(\bc_N - \bc_D)\}$.

\paragraph{Nonlinear Potential.}
However, it is possible that no such weight vector $\bw$ exists, {\it
i.e.}, the final convex polyhedron ${\cal P} = \bigcap_{i=1}^n {\cal
P}_i$ may be an empty set.  First, for a specific native protein $i$,
there may be severe restriction from some inequality constraints,
which makes ${\cal P}_i$ an empty set.  Some decoys are very difficult
to discriminate due to perhaps deficiency in protein representation.
In these cases, it is impossible to adjust the weight vector so the
native structure has a lower energy than the decoy.
Second,
even if a weight vector $\bw$ can be found for each native protein,
{\it i.e.}, $\bw$ is contained in a nonempty polyhedron, it is still
possible that the intersection of $n$ polyhedra is an empty set, {\it
i.e.}, no weight vector can be found that can stabilize all native
proteins against the decoys simultaneously.  Computationally, the
question whether a solution weight vector $\bw$ exists can be answered
unambiguously in polynomial time \cite{Karmarkar84}, and recent
studies using millions of decoys strongly suggest that when the number
of native protein structure reaches 300--400, no such weight vector
can be found \cite{Vendruscolo00_Proteins,TobiElber00_Proteins_1}. When
perfect discrimination is impossible, a technique that minimizes the
percentage of unsatisfied inequalities or the error rate was developed
in reference \cite{XiaLevitt00_JCP}.

A fundamental reason for this failure is that the functional form of
linear sum of pairwise interaction is too simplistic.  It has been
suggested that higher order interactions such as three-body or
four-body contacts should be incorporated
\cite{Betancourt99_PS,Munson97,Zheng97}. Functions with polynomial terms
using upto 6 degree of Chebyshev expansion has also been used to
represent pairwise interactions \cite{Fain02_PS}.

Here we propose an alternative approach.  At this time we still limit
ourselves to pairwise contact interactions.  We introduce a nonlinear
potential function analogous to the dual form of the linear function
in Equation (\ref{LinearEq}), which takes the following form:
\begin{equation}
H(f(\bs, \ba)) =  H(\bc) =
\sum_{D \in {\cal D}}\alpha_D \cdot K(\bc, \bc_D) -
\sum_{N \in {\cal N}}\alpha_N \cdot K(\bc, \bc_N)
\label{nonlinear}
\end{equation}
where $\alpha_D \ge 0$ and $\alpha_N \ge 0$ are parameters of the
potential function to be determined, and $\bc_D$ is a contact vector
of decoy $D$ in the set of decoys $\cal D = \{ (\bs_D, \ba_N )\}$,
$\bc_N$ a contact vector of native structure $N$ in the set of native
training proteins ${\cal N} = \{(\bs_N, \ba_N) \}$. The difference of
this functional form from linear function in Equation (\ref{LinearEq})
is that a kernel function $K(\bx, \by)$ replaces the linear term.  A
convenient kernel function $K$ is:
\[K(\bx, \by) =
e^{-||\bx -\by||^2/2\sigma^2}.
\]
Intuitively, the potential surface has smooth Gaussian hills of height
$\alpha_D$ centered on the location $\bc_D$ of decoy structure $D$,
and has smooth Gaussian cones of depth $\alpha_N$ centered on the
location $\bc_N$ of native structures $N$. Ideally, the value of the
potential function will be $-1$ for contact vectors $\bc_N$ of native
proteins, and will be $+1$ for contact vectors $\bc_D$ of decoys.

\paragraph{Optimal Nonlinear Potential.}

To obtain the nonlinear potential, our goal is to find a set of parameters
$\{\alpha_D, \alpha_N\}$ such that $ H(f(\bs_N, \ba_N))$ has energy
value close to $-1$ for native proteins, and the decoys have energy
values close to $+1$.  There are many different choices of
$\{\alpha_D, \alpha_N \}$.  We use an optimality criterion originally
developed in statistical learning theory
\cite{Vapnik95,Burges98,ScholkopfSmola02}.  First, we note that we
have implicitly mapped each structure and decoy from $\real^{210}$
through the kernel function of $K(\bx, \by) = e^{-||\bx
-\by||^2/2\sigma^2}$ to another space with dimension as high as tens
of millions.  Second, we then find the hyperplane of the largest
margin distance separating proteins and decoys in the space
transformed by the nonlinear kernel.  That is, we search for a
hyperplane with equal and maximal distance to the closest native
proteins and the closest decoys.  Such a hyperplane can be found by
obtaining the parameters $\{ \alpha_D \}$ and $\{ \alpha_N \}$ from
solving the following Lagrange dual form of quadratic programming
problem:
\begin{eqnarray*}
\mbox{Maximize } & \sum_{i\in  {\cal N} \cup {\cal D}}  
\alpha_i -
 \frac{1}{2}\sum_{i,j\in {{\cal N}\cup {\cal D}}}
y_i y_j \cdot \alpha_i \alpha_j \cdot e^{-||\bc_i - \bc_j||^2/2\sigma^2}
\\
\mbox{subject to} & 0\le \alpha_i \le C\\
\end{eqnarray*}
where $C$ is a regularizing constant that limits the influence of each
misclassified conformation 
\cite{VapChe64,VapChe74,Vapnik95,Burges98,ScholkopfSmola02}, and
$y_i =+1$ if $i$ is a native protein, and $y_i= -1$ if $i$ is a decoy.
These parameters lead to optimal classification of unseen test set
proteins against decoys \cite{VapChe64,VapChe74,Vapnik95,Burges98,ScholkopfSmola02}.

\paragraph{Modeling Sequence Design Potential.}

For protein sequence design, we assume that native sequence is more
stable on the native conformation than on a different structure of
another protein with low sequence identity.  We use the method of
gapless sequence threading to generate sequence decoys by mounting a
sequence fragment from a different protein of a larger size of an
unrelated structure to the native structure
\cite{MaiorovCrippen92_JMB}.  Because all native contacts are
retained, we find that such sequence decoys are quite challenging.

For protein design, we seek potential functions that allows the search
and identification of sequences most compatible with a specific given
coarse-grain three-dimensional structure.  We use a model analogous to
the Anfisen experiments.  We require that the native amino acid
sequence $\ba_N$ mounted on the native structure $\bs_N$ has the
lowest energy compared to a set of unrelated alternative sequences
${\cal D} = \{\bs_N, \ba_D\}$ mounted on the same native protein
structure $\bs_N$:
\[
H(f(\bs_N, \ba_N)) < H(f(\bs_N, \ba_D)) \quad \mbox{for all } \ba_D \in {\cal D}
\]
Equivalently, the native sequence will have the highest probability
to fit into the specified native structure.  This is the same
principle described in
\cite{ShakhGutin93_PNAS,Deutsch96_PRL,Li96_Science}.  Much work has been
done using linear
design function of sum of contact pairs in the form of
$H(f(\bs, \ba)) = H(\bc) = \bw \cdot \bc$
\cite{ShakhGutin93_PNAS,Deutsch96_PRL}.  The discussion of the two
geometric views of the weight vector of linear potential and the
optimality criterion also applies to the sequence design problem.

We now explore nonlinear potential function for sequence design using
the following functional form:
\begin{equation}
H(f(\bs, \ba)) =  H(\bc) =
\sum_{D \in \cal D}\alpha_D \cdot K(\bc, \bc_D) -
\sum_{N \in \cal N}\alpha_N \cdot K(\bc, \bc_N)
\label{nonlinear.seq}
\end{equation}
where $\alpha_D \ge 0$ and $\alpha_N \ge 0$ are coefficients to be
determined, and $\bc_D = f(\bs_N, \ba_D)$ is the contact vector of a
decoy sequence $\ba_D$ mounted on its native protein structure
$\bs_N$, and $\bc_N = f(\bs_N, \ba_N)$ is the contact vector of a
native sequence $\ba_N$ from the set of native training proteins $\cal
N$ mounted on the native structure $\bs_N$.  The only difference from
nonlinear folding potential of Equation (\ref{nonlinear}) is
that here $\cal D$ is a set of sequence decoys mounted on native
protein structures, rather than a set of structure decoys.  Again, we
use kernel function $K(\bx, \by) = e^{-||\bx -\by||^2/2\sigma^2}$.

The optimal parameters $\{\alpha_N\}$ and $\{ \alpha_D \}$ are
obtained similarly by solving the following Lagrange dual convex
quadratic programming problem:
\begin{eqnarray*}
&\mbox{Maximize }& \sum_{i\in  {{\cal N} \cup {\cal D}} } \alpha_i - \frac{1}{2}
\sum_{i,j \in {{\cal N} \cup {\cal D}}} y_i y_j \cdot \alpha_i \alpha_j e^{-||\bc_i - \bc_j||^2/2\sigma^2}
\\
&\mbox{subject to}& 0\le \alpha_i \le C
\end{eqnarray*}
where $C$ is a regularizing constant \cite{Vapnik95,Burges98},
and $y_i = +1$ if $i$ is a native protein, and $y_i = -1$ if $i$ is
a decoy.

\section{Computational Methods}
\paragraph{Alpha Contact Maps.}
Because protein molecules are formed by thousands of atoms, their
shapes are complex.  In this study we use the count vector of pairwise
contact interactions derived from the edge simplices of the alpha
shape of a protein structure.  Edge simplices in the alpha shape
represent nearest neighbor interactions that are in physical contacts.
They encode precisely the same contact information as a subset of the
edges in the Voronoi diagram of the protein molecule.  These Voronoi
edges are shared by two interacting atoms from different residues, but
intersect with the body of the molecule modeled as the union of atom
balls.  We refer to references \cite{Edels95_DCG,Liang98a_Proteins}
for further theoretical and computational details.

\paragraph{Generating Structure Decoys and Sequence Decoys by Threading.}
      \begin{figure}[t]
      \centerline{\epsfig{figure=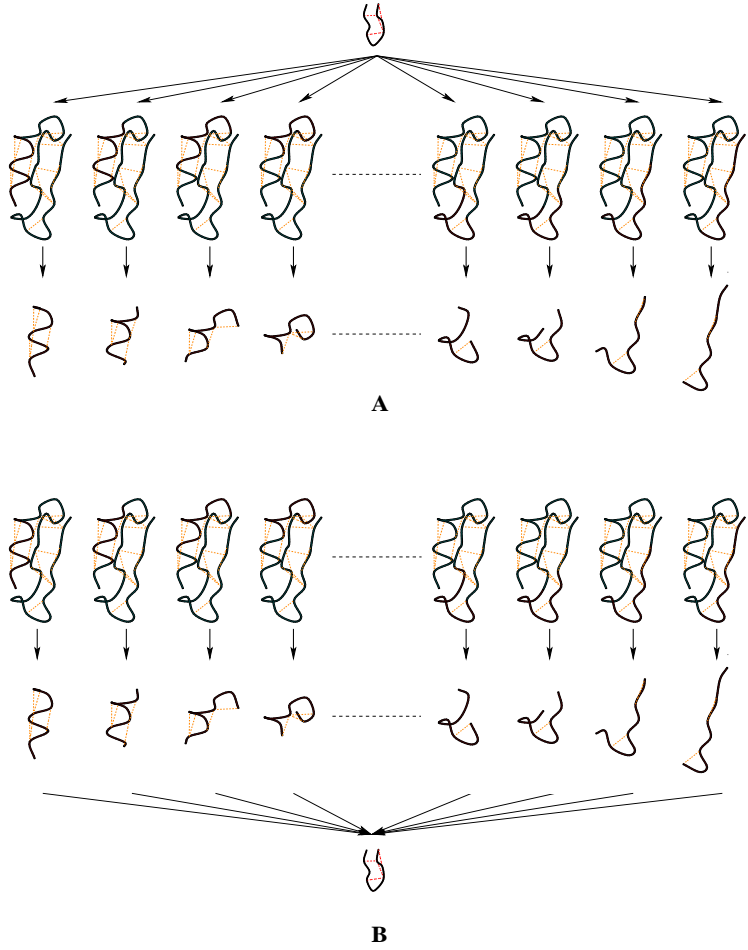,width=4in}}
\caption{\sf
Decoy generation by gapless threading.
(a). Structure decoys can be
generated by threading the sequence of a smaller protein to the
structure of an unrelated larger protein.
(b). Sequence decoys can be generated by threading the sequence of a
larger protein to the structure of an unrelated smaller protein.
}
      \label{Fig:Thread.structure}
\end{figure}

Maiorov and Crippen used the gapless threading method for
generating a large number of structure decoys
\cite{MaiorovCrippen92_JMB}. In this method, the sequence of a smaller
protein $\ba_N$ is threaded through the structure of an unrelated
larger protein and takes the conformation $\bs_D$ of a fragment of the
larger protein \cite{MaiorovCrippen92_JMB}.  Along the way, the
sequence of the smaller protein can take the conformations of many
fragments of the larger protein, each provides a structure decoy.
With this approach, we generate for each native protein $\{ (\bs_N,
\ba_N) \}$ a set of structure decoys $\{ (\bs_D, \ba_N) \}$
(Fig~\ref{Fig:Thread.structure}a).

We can generate sequence decoys in an analogous way, as already
suggested in \cite{Jones92_Nature,Munson97}.  In this case, we
thread the sequence of a larger protein through the structure of a
smaller protein, and obtain sequence decoys by mounting a fragment of
the native sequence from the large protein to the full structure of
the small protein.  We therefore have for each native protein $(\bs_N,
\ba_N)$ a set of sequence decoys $(\bs_N, \ba_D)$
(Fig~\ref{Fig:Thread.structure}b).

\paragraph{Protein Data.}
Following reference \cite{Vendruscolo00_Proteins}, we use protein structures
contained in the {\sc Whatif} database  \cite{WHATIF} in this study.
{\sc Whatif} database contains a representative set of sequence-unique
protein structures generated from X-ray crystallography.  Structures
selected for this study all have pairwise sequence identity $< 30\%$,
R-factor $<0.21$, and resolution $<2.1$.  {\sc Whatif} database
contains less structures than {\sc Pdbselect} because the R-factor and
resolution criteria are more stringent \cite{WHATIF}.  Nevertheless,
it provides a good representative set of currently all known protein
structures.

We use a list of 456 proteins compiled from the 1998 release ({\sc
Whatif98}) of the {\sc Whatif} database \cite{Vendruscolo00_Proteins},
which was kindly provided by Dr.\ Vendruscolo.  There are 192 proteins
with multiple chains in this dataset.  Some of them have extensive
interchain contacts.  For these proteins, it is possible that their
conformations may be different if there are no interchain contacts
present.  We use the criterion of {\it Contact Ratio} to remove
proteins that have extensive interchain contacts.  Contact Ratio is
defined here as the number of interchain contacts divided by the total
number of contacts a chain makes. For example, protein {\tt 1ept} has
four chains A,B,C, and D. The intra chain contact number of chain B is
397.  Contacts between chain A and chain B is 178, between B and C is
220, between B and other heteroatoms is 11.  The Contact Ratio of
chain B is therefore $ (178+220+11)/ (397+178+220+11) = 51 \% $.
Thirteen protein chains are removed because they all have Contact
Ratio $ > 30 \% $.  We further remove three proteins because each has
$>10\% $ of residues missing with no coordinates in the Protein Data
Bank file.  The remaining set of 440 proteins are then used as
training set for developing both folding and design potential
functions.  Using the sequence and structure threading method
described earlier, we generated a set of 14 080 766 sequence decoys and
a set of structure decoys of the same size.

\paragraph{Learning Linear Potential.}
For comparison, we develop optimal linear potential following the method
and computational procedure described in reference\cite{TobiElber00_Proteins_1}. We apply the interior
point method as implemented in BPMD by M\'esz\'aros \cite{Meszaros96_CMA} 
to
search for a weight vector $\bw$.  We use two different
optimization criteria as described in reference
\cite{TobiElber00_Proteins_1}.  The first is:
\begin{eqnarray*}
\mbox{Identify} & \bw 
\\
\mbox{subject to} & \bw \cdot (\bc_N - \bc_D) < \epsilon  \quad \mbox{ and } |\bw_i| \le 10\\
\end{eqnarray*}
where $\bw_i$ denotes the $i$-th component of weight vector $\bw$, and
$\epsilon = 1 \times 10^{-6}$.  Let $ {\cal C} = \{ c_N - c_D \}$, and
$|\cal C|$ the number of decoys.  The second optimization criterion
is:
\begin{eqnarray*}
\mbox{Minimize }  & 
 \min \frac{1}{|{\cal C}|} \sum \left( \bw \cdot (c_N - c_D)\right)^2 - 
                         \left[ \frac{1}{|{\cal C}|}\sum  \left(\bw \cdot (\bc_N -\bc_D)\right) \right]^2 
 \\
\mbox{subject to} & \bw \cdot (\bc_N - \bc_D) < \epsilon \\
\end{eqnarray*}

\paragraph{Learning Nonlinear Kernel Potential.}
We use {\sc SVMlight} ({\tt http://svmlight.joachims.org/})
\cite{Joachims99} with Gaussian kernels and a training set of 440
native proteins plus 14 080 766 decoys to obtain the optimized
parameter $\{\alpha_N, \alpha_D\}$.
The regularization constant $C$  takes  default value, which is
estimated from the training set ${{\cal N} \cup {\cal D}}$ as implemented in {\sc SVMlight}:
\begin{equation} 
C =  |{\cal N} \cup {\cal D}|^2 / \left [ \sum_{\bx \in {\cal N} \cup {\cal D}} \sqrt{K(\bx,
\bx)-2 \cdot K(\bx,{\bf 0}) + K({\bf 0, 0})} \right ]
^2.
\end{equation}

Since we cannot load all 14 millions decoys into computer memory
simultaneously, we use three heuristic strategies for training.  In
the first method, the total 14 080 766 decoys are divided randomly
into 34 subsets.  The training process starts with all native proteins
and the first decoy set.  After the training process converges, we
select the decoy structures that have $\alpha_i \ne 0$ (these decoys
are called support vectors \cite{Vapnik95,Burges98,ScholkopfSmola02}).
From the second iteration on, we combined the native proteins, the new
decoy set, and all decoys that appeared as support vectors in any of
the previous training steps.  These form the new training set for the
next iteration of learning.  This process is continued until all
decoys have been exhausted in training.

In the second method, each of the 34 decoy subsets was combined with
the native set in turn to form a training set.  The decoy support
vectors are then selected from each of the 34 training processes. All
these decoy support vectors are combined, along with native proteins
for a final training process.  The results are taken as the optimized
protein potential.

The third method is similar to the procedure reported in
\cite{TobiElber00_Proteins_1}.  We first randomly selected a subset of
decoys that fits into the computer memory.  For example, we pick every
51st decoy from the list of 14 million decoys.  This leads to an
initial training set of 276 095 decoys and 440 native proteins. An
initial protein potential is then obtained after learning.  Next the
energies for all 14 million decoys and all 440 native proteins are
evaluated.  Three decoy sets were collected based on the evaluation
results: the first set of decoys contains the violating decoys which
have lower energy than the native structures; the second set contains
decoys with the lowest absolute energy, and the third set contains
decoy support vectors identified in previous training process.  The
union of these three subsets of decoys are then combined with the 440
native protein as the training set for the next iteration of
learning. This process is repeated until the energy difference to
native protein for all decoys are greater than 0.0.  Using this
strategy, the number of iterations typically is between 2 and 10.
During the training process, we set the cost factor $j$ in {\sc
SVMlight} to 120, which is the factor by which training errors on
native proteins outweighs errors on decoys.

The value of $\sigma^2$ for the Gaussian kernel $K(\bx, \by) =
e^{-||\bx -\by||^2/2\sigma^2}$ is chosen by experimentation.  We find
that if the value of $\sigma^2$ is too large, no $\{\alpha_N,
\alpha_D\}$ can be found that can perfectly classifies the 440
training proteins and their decoys, {\it i.e.}, the problem is
unlearnable. If the value of $\sigma^2$ is too small, the performance
in blind-test will deteriorate because of overfitting.  For the 440
native proteins and the 14 080 766 sequence decoys, the value of
$\sigma^2$ is chosen between 250.0 (for training method 1) and 138.9
(for training method 3).  The final folding potential is obtained with
$\sigma^2= 227.3$ and the final design potential is obtained with
$\sigma^2 = 138.9$, both derived using the third training method.

\section{Results}

\paragraph{Linear Folding Potentials from Structure Decoys.}
Using perceptron learning and a large number of decoys generated by
gapless threading, Vendruscolo {\it et al\/} showed that if the number
of native proteins exceeds 270, it is impossible to find parameters
$\bw$ for potential function $H(\bs, \ba) = \bw \cdot \bc$, such that
all native structures have lower energies than decoys
\cite{Vendruscolo00_Proteins}.  That is, no $\bw$ can be found such
that $ \bw \cdot \bc_N < \bw \cdot \bc_D $ holds for all decoys.

Tobi {\it et al\/} showed that pairwise contact potential $H(\bs, \ba)
= \bw \cdot \bc$ can be found to distinguish a different set of 572
proteins from a set of 28 213 009 structure decoys generated by
gapless threading \cite{TobiElber00_Proteins_1}.  In this study, two
residues are defined to be in contact if the geometric centers of
their side chains are at a distance between 2.0 \AA\ and 6.4 \AA.  To
search for the optimal weight vector $\bw$, the authors used linear
programming solver based on interior point method as implemented in BPMD
by M\'esz\'aros \cite{Meszaros96_CMA}.

We succeeded in reproducing the results of Tobi {\it et al}
\cite{TobiElber00_Proteins_1} and found a $\bw$ vector that enables
perfect discrimination of the same set of 572 proteins used in
\cite{TobiElber00_Proteins_1} against 28 261 307 structure decoys we
generated by gapless threading.  We used the same contact definition
based on Euclidean distance between geometric centers of side chains,
as well as the same optimization criterion described earlier in the
Methods section.  The values of the 210 pairwise contact potentials
(data not shown) are similar although not identical to the values
listed in \cite{TobiElber00_Proteins_1}.

\paragraph{Linear Design Potentials from Sequence Decoys.}

Structure decoys generated by gapless threading are obtained by taking
a fragment of the structure of a large protein such that it contains
exactly the same number of amino acid residues as the native protein.
However, contacts in the resulting fragment structure often contain
only a fraction of the total contacts in
native protein.

      \begin{figure}[tbh]
      \centerline{\epsfig{figure=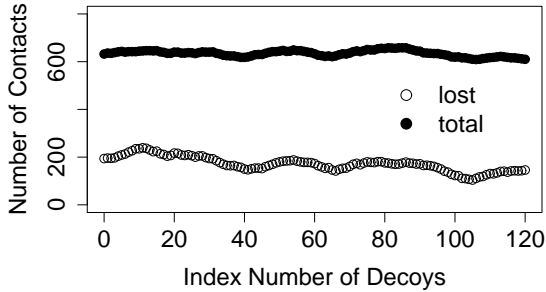,width=3in}}
\caption{\sf
Structure decoys generated by
threading have many contacts lost. There is 185 residues in protein
{\tt 1531}, and 305 residues in {\tt 1abe}.  There are 121 decoys when
the sequence of {\tt 1531} is threaded on the segments of the
structure of {\tt 1abe}.  Between 1/3 -- 1/4 of contacts are lost in
these decoys. Structure decoys generated by threading are therefore not
very challenging.
}
      \label{Fig:Thread.diff}
\end{figure}

Design decoys are generated by taking a fragment sequence of a larger
protein and thread it onto the native conformation of a smaller
protein.  As a result, although the contact count vector $\bc$ may be
very different, the number of contacts in the design decoy is exactly
the same as that in native protein.  For decoys generated by gapless
threading, design decoys therefore are far more challenge to
discriminate then structure decoys (Fig~\ref{Fig:Thread.diff}).

Our experimentation confirms this observation.  After generating 14
080 766 sequence design decoys for the 440 proteins in the training
set, we apply the same interior point method to search for an optimal
$\bw$ that can discriminate native sequences from decoy
sequences. That is, we search for parameters $\bw$ for $H(\bs, \ba) =
\bw \cdot \bc$, such that $ \bw \cdot \bc_N < \bw \cdot \bc_D$ for all
sequences.  However, we fail to find a feasible solution for the
weight vector $\bw$.  This indicates that no $\bw$ exists capable of
discriminating perfectly 440 native sequences from the 14 million
decoy sequences.  We repeated the same experiment using the set of 572
native proteins from reference \cite{TobiElber00_Proteins_1} and 28 261
307 sequence decoys.  The result is also negative.

\paragraph{Learning Nonlinear Kernel potential.}
To overcome the problems associated with linear potentials, we use the
same set of 440 native proteins and 14 million decoys to obtain
nonlinear kernel folding and design potentials.  In both cases, we
succeeded in finding a function in the form of Equation
(\ref{nonlinear}) that can discriminate all 440 native
proteins from 14 million decoys.

\begin{table}[htb]
\begin{small}
\begin{center}
\vspace{.3in}
\begin{tabular}{|c|c|c|c|c|c|}
\hline
\multicolumn{2}{|c|}{} &\multicolumn{3}{|c|}{Design Potential} & Folding Potential\\ \cline{3-6}
\multicolumn{2}{|c|}{} & Strategy 1   & Strategy 2 & Strategy 3& \\
\multicolumn{2}{|c|}{} & $\sigma^2=250.0$ & $\sigma^2=227.3$ & $\sigma^2 = 138.9$ & $\sigma^2 = 227.3$\\ \hline

Num.\ of        & Natives & 260 & 258 & 347&268 \\ \cline{2-6}
Support Vectors & Decoys  & 2457 & 2877 & 4709& 2560\\ \hline
Range of        & Natives & $1.070 \sim 0.9996$ & $0.9993 \sim 5.762$ & $0.9991 \sim 3.314$ & $0.9990 \sim  4.485$ \\ \cline{2-6}

Energy Values   & Decoys  & $-9.495 \sim 0.7979$ & $-9.485 \sim 0.9339$ & $-6.655 \sim 0.9882$ & $-8.379 \sim 1.039$\\ \hline
\multicolumn{2}{|c|}{Range of Smallest Energy Gap } & $0.2025 \sim 14.56 $ & $0.06624 \sim  14.55$ & $0.01226 \sim 9.237 $ & $0.0610 \sim 11.36 $ \\
\hline
\end{tabular}
\caption{\sf
 Details of learning nonlinear kernel
folding and design potentials.  The number of native proteins and
decoys with non-zero $\alpha_i$ entering the potential function is
listed.  These native proteins and decoys are called support vectors.
The range of the energy values of natives and decoys are also listed,
as well as the range of the energy gaps between the native protein and
the decoy with the lowest energy.
}
\label{tab:learning}
\end{center}
\end{small}
\end{table}

Unlike statistical potentials where each native protein structure in
the database contribute to the empirical potential, only a subset of
native structures contribute and have $\alpha_N \ne 0$.  In addition,
a small fraction of decoys also contribute to the potential function.
Table~\ref{tab:learning} list the details of the potential, including
the numbers of native proteins and decoys that participate in Equation
(\ref{nonlinear}).  These number represent about $ 60\% $ of native
proteins and $<0.1\%$ of decoys from the original training data.

\paragraph{Discrimination Tests for Folding Potential.}
Blind test in discriminating native proteins from decoys using an
independent test set is essential to assess the effectiveness of
folding potentials.  To construct a test set, we first take the
entries in {\sc Whatif99} database that are not present in {\sc
Whatif98}.  After eliminating proteins with chain length less than 46
residues, we obtain a set of 201 proteins.  These proteins all have
$<$ 30\% sequence identities with any other sequence in either the
training set or the test set proteins.  Since 139 of the 201 test
proteins have multiple chains, we use the same criteria applied in
training set selection to exclude 7 proteins with $>30\%$ Contact
Ratio or with $>10\%$ residues missing coordinates in the PDB files.
This leaves a smaller set of test proteins of 194 proteins.  Using
gapless threading, we generate a sets of 3 096 019 structure decoys
from the set of 201 proteins. This is a superset of the decoy set
generated using 194 proteins.

For comparison, we also test the discrimination results of the optimal
linear potential taken as reported in reference
\cite{TobiElber00_Proteins_1}, as well as the statistical potential
developed by Miyazawa and Jernigan.  Here we use the contact
definition reported in \cite{TobiElber00_Proteins_1}, that is, two
residues are declared to be in contact if the geometric centers of
their side chains are within a distance of 2.0 -- 6.4 \AA.

\vspace*{.3in}
\begin{table}[thb]
\begin{center}
\begin{tabular}{|c|c|c|}
\hline
    & { Misclassified Natives} & {Misclassified Natives} \\ \hline \hline
{ Kernel Folding Potential}    &  3/194 & 8/201  \\ \hline
Tobi \& Elber$^\&$              &  7/194 & 13/201 \\ \hline
Miyazawa \& Jernigan            &  85/194& 92/201 \\ \hline
\hline
{ Kernel Design Potential}      &  7/194 &13/201  \\ \hline
\end{tabular}\\
\caption{\sf  The number of misclassified
protein structures using nonlinear kernel folding potential, optimal
linear potential taken as reported in \cite{TobiElber00_Proteins_1},
and Miyazawa-Jernigan statistical potential \cite{Miyazawa96_JMB}
among the test set of 194 proteins and the set of 201 proteins. The latter
include proteins with more than 30\% interchain contacts and proteins
with $>10\%$ missing coordinates.  We also list performance of kernel
design potential for structure recognition.
}
\label{tab:result.fold}
\end{center}
\end{table}

To test nonlinear folding potential functions for discriminating
native proteins from structure decoys in both the 194 and the 201 test
sets, we take the structure $\bs$ from the conformation-sequence pair
$(\bs, \ba_N)$ with the lowest energy as the predicted structure of
the native sequence.  If it is not the native structure $\bs_N$, the
discrimination failed and the folding potential does not work for this
protein.  The results of discriminating native structures using nonlinear
folding potential are summarized in Table~\ref{tab:result.fold}.
There are 3 and 8 misclassified native structures for the 194 set and
201 set, respectively. These correspond to a failure rate of 1.5\% and
4.0\%, respectively.  The optimal nonlinear kernel folding potential
performs better than the optimal linear potential based on calculation
using potential values taken as reported in reference
\cite{TobiElber00_Proteins_1} (failure rates 3.6\% and 6.5\% for the
194 set and 201 set, respectively).  Consistent with previous reports
\cite{Clementi98_PRL}, statistical potential has about $43.8\%$ (81
out of 194) and $43.2\%$ (87 out of 201) failure rates for the 194 set
and the 201 set, respectively.

\paragraph{Discrimination Tests for Design Potential.}
We use the same 194 set and 201 set of natives proteins and generate a
set of 3 096 019 sequence decoys for testing the design potential.  We
take the sequence $\ba$ from the conformation-sequence pair $(\bs_N,
\ba)$ for a protein with the lowest energy as the predicted sequence.
If it is not the native sequence $\ba_N$, the discrimination failed
and the design potential does not work for this protein.

Sequence decoys obtained by gapless threading are quite challenging,
since all native contacts of the protein structures are maintained.
No linear design potential function can be found using linear
programming method that would succeed in the challenging task of
identifying all 440 native sequences in the training set.

\begin{table}[tb]
\begin{center}
\label{tab:result.design}
\vspace*{.3in}
\begin{tabular}{|c|c|c|}
\hline
    & { Misclassified Natives} & {Misclassified Natives} \\ \hline \hline
{ Kernel Design Potential$^*$}  &  14/194 & 20/201 \\ \hline
Tobi \& Elber$^\&$              &  37/194 & 44/201 \\ \hline
Miyazawa \& Jernigan            &  81/194 & 87/201 \\ \hline \hline
{ Kernel Folding Potential}     &  24/194 & 30/201 \\ \hline
\end{tabular}\\
\caption{\sf
 The number of misclassified
protein sequences using nonlinear kernel design potential, optimal
linear potential taken as reported in \cite{TobiElber00_Proteins_1},
and Miyazawa-Jernigan statistical potential \cite{Miyazawa96_JMB}
among the set of 194 proteins and the set of 201 proteins. The latter
include proteins with more than 30\% interchain contacts and proteins
with $>10\%$ missing coordinates.  The nonlinear kernel design
potential is the only function that succeeded in perfect discrimination
of the 440 native sequences from a set of 14 million sequence decoys.
}
\end{center}
\end{table}

We succeeded in obtaining nonlinear design potential capable of
discriminating all of the 440 native sequences
(Table~\ref{tab:result.design}).  It also works well in the test set.
It succeeded in correctly identifying 92.8\% (180 out of 194) of
native sequences in the independent test set.  This compares favorably
with results obtained using optimal linear folding potential taken as
reported in \cite{TobiElber00_Proteins_1}, which succeeded in
identifying 80.9\% (157 out of 194) of the test set, and the
Miyazawa-Jernigan statistical potential (success rate 58.2\%, 113 out
of 194).

\paragraph{Discrimination Test Using Challenging Decoys.}

\begin{figure}[t]
      \centerline{\epsfig{figure=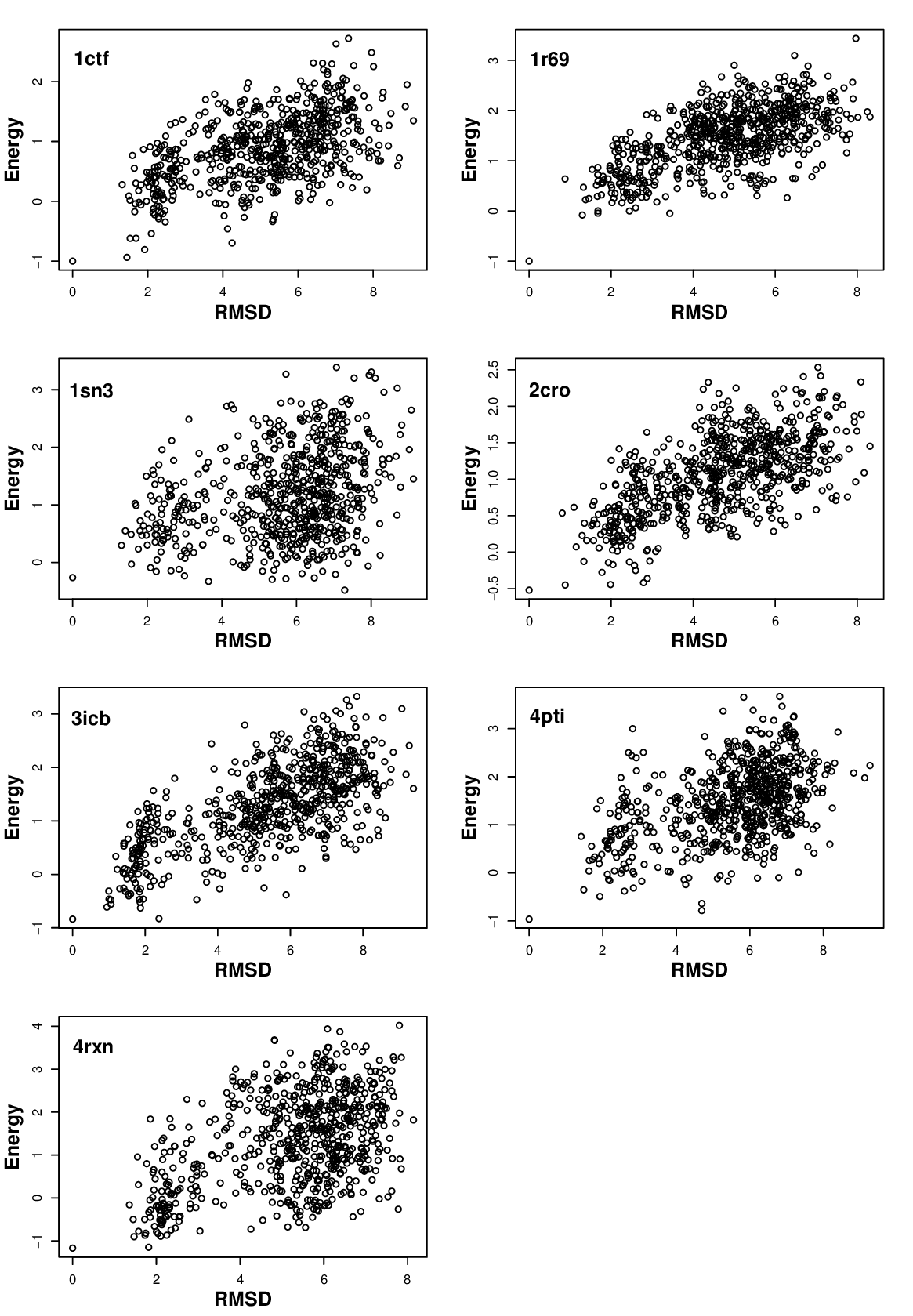,width=4in}}
\caption{\sf
 The energy values of decoys and native
proteins in the {\sc 4state\_reduced}  set by nonlinear design
potentials and their correlation with the rmsd values to the native
structures.
}
      \label{Fig:4state}
\end{figure}

For the protein potentials derived with simple decoys generated by
gapless threading, a more challenging test is to discriminate native
proteins from an ensemble of explicitly generated three dimensional
decoy structures with a significant number of near-native
conformations \cite{ParkLevitt96_JMB,SamudralaMoult98_JMB}.  Here we
evaluate the performance of nonlinear folding and design potential
using three decoy sets from the database ``{\sc Decoys 'R' Us}''
\cite{SamudralaLevitt00_PS}: the {\sc 4state\_reduced} set,  the {\sc
Lattice\_ssfit} set, and the {\sc lmsd} set.
We compare our results in performance with results reported in
literature using optimal linear potential \cite{TobiElber00_Proteins}
and statistical potential \cite{Miyazawa96_JMB}
(Table~\ref{tab:4state}).  For the {\sc 4state\_reduced} set of
decoys, nonlinear design potential has the best performance in terms
of identifying the native structure.  The only misclassified protein
{\tt 1sn3} has three disulfide bonds, which are not modeled explicitly
in the protein description of a vector in $\real^{210}$.  The
correlation of root mean square distance of conformations to the
native structure and energy value in the {\sc 4state} set are shown in
Fig~\ref{Fig:4state}.  For proteins {\tt 1r69, 2cro, 3icb}, and {\tt
1ctf}, the correlation coefficient between the rmsd values and the
energy values are good.  We note that the nonlinear potential is
obtained by learning from training decoys that are obtained by gapless
threading.  It is likely that nonlinear kernel potential can be
further improved if more realistic structure decoys are included in
training.

\begin{table}
\begin{small}
\begin{center}
\label{tab:4state}
\vspace*{-.06in}
\begin{tabular}{cccccc}
\multicolumn{6}{l}{1. {\sf 4state\_reduced} } \\ \hline
Protein & \# of decoys &  KFP & KDP  &   MJ   &   TE-13 \\ \hline
1ctf   & 631 &  1/3.31(0.51) &       1/3.26(0.53)   &  1/3.73 & 1/4.20    \\
1r69   & 676 &  1/3.52(0.58) &       1/4.04(0.62)   &  1/4.11 & 1/4.06    \\
1sn3   & 661 &  3/2.44(0.35) &       6/1.91(0.30)   &  2/3.17 & 6/2.70    \\
2cro   & 675 &  8/2.24(0.56 &       1/2.93(0.63)    &  1/4.29 & 1/3.48    \\
3icb   & 654 &  1/2.64(0.71) &       1/2.73(0.69)  &         &           \\
4pti   & 688 &  1/3.85(0.40) &       1/3.19(0.47)   &  3/3.16 & 7/2.43    \\
4rxn   & 678 &  1/2.71(0.49) &       1/2.29(0.48)   &  1/3.09 & 16/1.97   \\
\hline
\multicolumn{6}{l}{2. {\sf lattice\_ssfit}} \\ \hline
Protein & \# of decoys & KFP &  KDP &    MJ  & TE-13 \\ \hline
1beo    &   2001      & 3/3.04  & 10/2.61 &         &          \\
1ctf    &   2001      & 1/4.65  & 1/4.73  &  1/5.35 &  1/6.17  \\
1dkt    &   2001      & 2/3.42  & 3/3.12  &  32/2.41&  2/3.92  \\
1fca    &   2001      & 8/2.89  & 21/2.67 &  5/3.40 &  36/2.25 \\
1nkl    &   2001      & 1/3.68  & 1/3.81  &  1/5.09 &  1/4.51  \\
1pgb    &   2001      & 1/4.67  & 1/4.40  &  3/3.78 &  1/4.13  \\
1trl    &   2001      & 32/2.34 & 175/1.34&  4/2.91 &  1/3.63  \\
4icb    &   2001      & 1/4.52  & 1/4.62  &         &     \\ \hline

\multicolumn{6}{l}{3. {\sf 1msd}} \\ \hline
Protein & \# of decoys & KFP &  KDP &    MJ  & TE-13 \\ \hline
1b0n-B  &  498  &  45/1.33   &  129/0.61  &   &  \\
1bba    &  501  &  6/-3.63   &  501/-2.97 &  &  \\
1ctf    &  498  &  1/3.38    &  1/3.48    & 1/3.86   & 1/4.13 \\
1dtk    &  216  &  154/0.63  &  178/-1.13 & 13/1.71  & 5/1.88 \\
1fc2    &  510  &  501/-4.00 &  499/-2.90 & 501/-6.24& 14/2.04  \\
1igd    &  510  &  1/4.05    &  1/3.85    & 1/3.25   & 2/3.11 \\
1shf-A  &  438  &  2/2.64    &  2/2.44    & 11/2.01  & 1/4.13 \\
2cro    &  501  &  1/2.77    &  1/5.13    & 1/5.07   & 1/3.96 \\
2ovo    &  348  &  1/1.21    &  67/0.93   & 2/3.25  &  1/3.62\\
4pti    &  344  &  7/1.57    &  47/0.97   &  &  \\
\hline
\end{tabular}
\end{center}
\caption{\sf
 Results of discrimination of native
structures from decoys using nonlinear kernel potentials.  The decoy
sets include the {\sc 4state\_reduced} set,
the {\sc Lattice\_ssfit} set, and the {\sc lmsd} set
\cite{SamudralaLevitt00_PS}.
The energy rank of the native structure and its $z$-score are listed.
The correlation coefficient $R$ is also listed in parenthesis for the
{\sc 4state\_reduced} set.  KFP stands for kernel folding potential,
and KDP stands for kernel design potential. TE-13 potential is linear
distance based potential optimized by linear programming, taken as
reported in \cite{TobiElber00_Proteins}, and MJ potential is
statistical potential as reported in \cite{Miyazawa96_JMB}.  Results
for TE-13 potential and Miyazawa-Jernigan potential are taken from
Table II of \cite{TobiElber00_Proteins}.
}
\end{small}
\end{table}

\paragraph{Nonlinear Potential Function for Folding and Design.}
      \begin{figure}[t]
      \centerline{\epsfig{figure=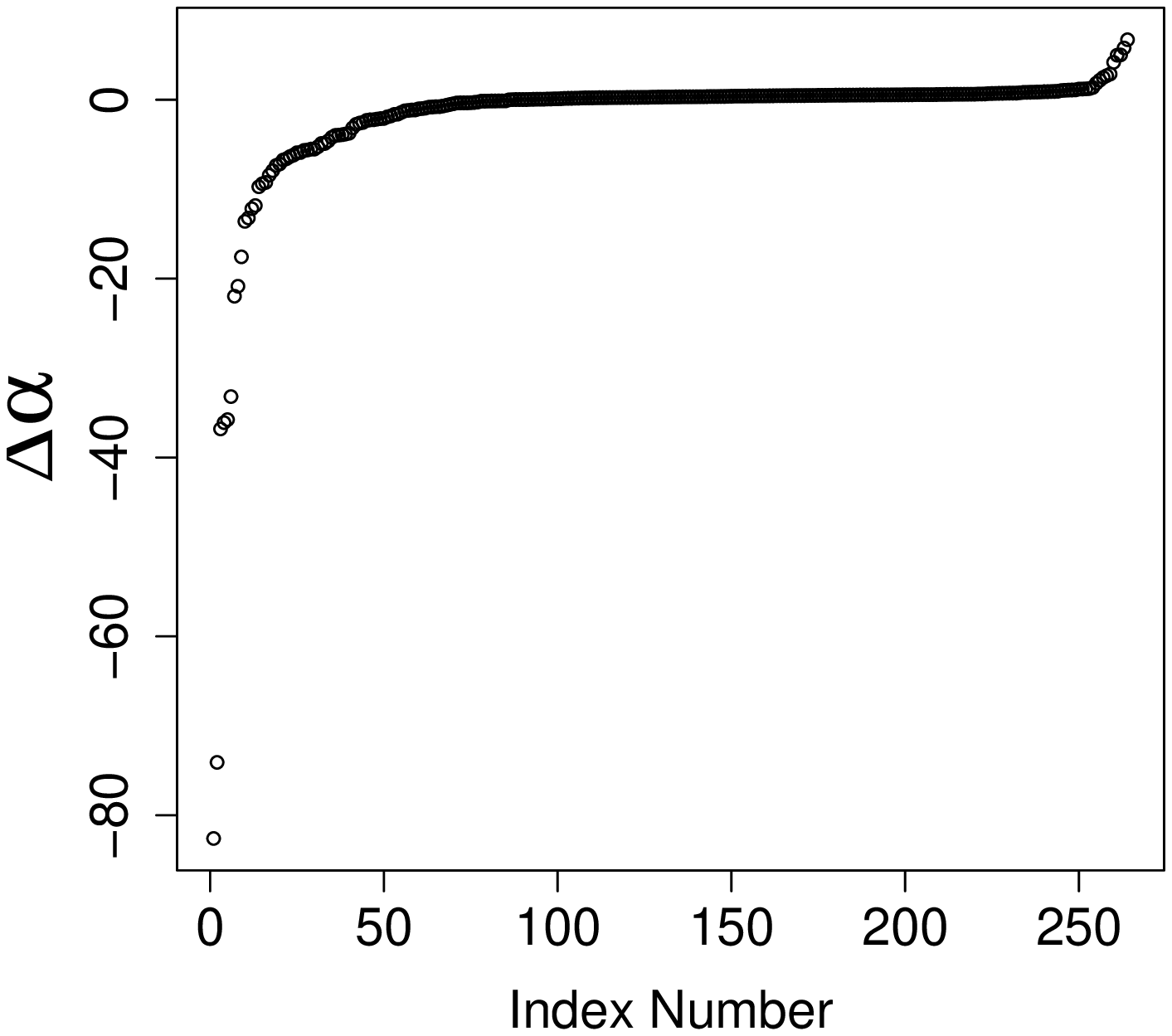,width=3in}}
\caption{\sf
 The difference in contribution to
the potential function for native protein structures that participate
in both folding and design potentials.  They are sorted by $\Delta
\alpha = \alpha_{\mbox{design}} - \alpha_{\mbox{folding}}$.  The
majority of them have $\Delta \alpha$ close to 0.
}
      \label{Fig:alphaDiff}
\end{figure}

\begin{figure}[tb]
 \centerline{\epsfig{figure=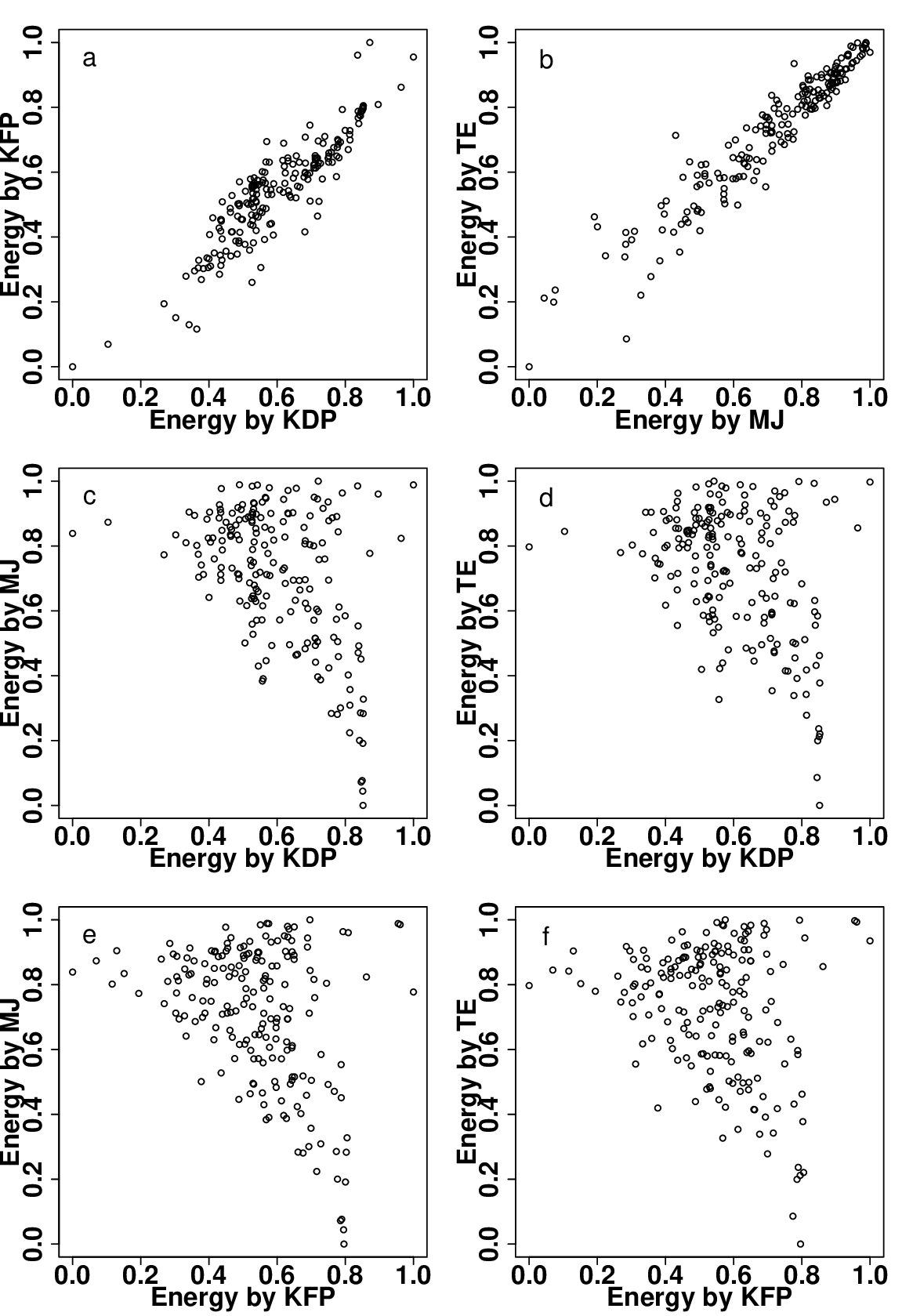,width=3in}}
\caption{\sf
Comparison of nonlinear kernel
folding potential (KFP) and kernel design potentials (KDP) with
Tobi-Elber (TE) optimal linear potential and Miyazawa-Jernigan (MJ)
statistical potential.  The energy values of the 194 proteins in the
test set are calculated and scaled to 0.0 -- 1.0.  (a). The energy values by KFP and by KDP for
the 194 proteins are highly correlated.  The correlation coefficient
is $R = 90$.   The energy values by (b) MJ and TE are also highly correlated.
The correlation coefficient is $R = 0.95$.
The energy values by (c) MJ and KDP, (d) by TE and KDP,
(e) by MJ and KFP, (f) by TE and KFP are all poorly correlated.  This
suggests that both the kernel folding and design potential functions
are different from MJ and TE potentials.
}
      \label{Fig:testSetenergies}
\end{figure}

Because nonlinear potential is expressed as a kernel expansion of
native protein and decoys, structure decoys and sequence decoys in
general lead to different protein functions.  For example, the contact
count vectors $\bc$ can be very different for a sequence decoy of a
protein and a structure decoy of the same protein.  The potential
surface defined by the folding potential and the design potential
therefore may be different.  There are 268 out of 440 native proteins
participating in folding potential function with $\alpha$ value ranging
from 0.01 - 28.04, of which 185 ($65\%$) are between 0.01 and 2.00.  There
are 347 out of 440 native proteins participating in design potential
function, with $\alpha$ value ranging from 0.02 to 110.64,  of
which 269 ($77\%$) are between 0.02 and 2.00.  Therefore, the majority of
the native proteins have similar $\alpha$ values for
both folding and design potentials. 
Fig~\ref{Fig:alphaDiff} shows the difference $\Delta \alpha_i$ of the
coefficient $\alpha_i$ for protein $i$ appearing in both folding
potential and design potential. For the majority of the native
proteins, $\Delta \alpha_i $ values are small.  That is, most native
proteins contribute similarly in design potential and in folding
potential.  This is expected, because the main differences between the
two potentials are due to differences in decoy sets.  The native
proteins that contribute most differently to folding potential and
design potential are small proteins.  For example, there are 34 native
proteins whose $ |\Delta \alpha| >5.0 $, and they all come from the
top 100 smallest protein chains.  Table~\ref{tab:highest_alpha} listed
20 proteins with highest $\alpha$ values for both kernel folding and
design potentials.  They are all small proteins and there are 15 out
of the 20 proteins that appear both in folding and design potentials.
It is possible that the energy values by kernel folding potential and
by kernel design potential may be similar for many structure-sequence
pairs $(\bs, \ba)$.  Figure~\ref{Fig:testSetenergies}a shows that the
test 194 proteins have similar energy values by the kernel folding and
kernel design potentials.

\begin{table}[htb]
\begin{small}
\begin{center}
\vspace*{.3in}
 \begin{tabular}{l|lccc|lccc}\hline
&\multicolumn{4}{c|}{Kernel Design Potential}&\multicolumn{4}{c}{Kernel Folding Potential} \\ \cline{2-9}
 Index  & pdb & $\alpha$  & Class & Number &   pdb & $\alpha$ & Class & Number\\
   &  &value  & & of resides &    &value  &  &  of resides  \\ \hline
1 & 1rop.a.pdb & 28.04 & a & 56 & 1rop.a.pdb & 110.64 & a & 56 \\
2 & 1vie.pdb & 21.43 & b & 60 & 1tgs.i.pdb & 91.76 & g & 56\\
3 & 1igd.pdb & 20.75 & d & 61 & 2spc.a.pdb & 56.93 & a & 107\\
4 & 2spc.a.pdb & 20.12 & a & 107 & 2ovo.pdb & 46.34 & g & 56\\
5 & 1abo.a.pdb & 20.08 & b & 58 & 1isu.a.pdb & 43.49 & g & 62\\
6 & 1ail.pdb & 19.92 & a & 70 & 6ins.e.pdb & 42.31 & g & 50\\
7 & 1mjc.pdb & 19.75 & b & 69 & 1ail.pdb & 40.77 & a & 70\\
8 & 2utg.a.pdb & 18.57 & a & 70 & 2utg.a.pdb & 40.54 & a & 70\\
9 & 1tgs.i.pdb & 17.67 & g & 56 & 1vie.pdb & 35.03 & b & 60\\
10 & 1tig.pdb & 11.52 & d & 88 & 1igd.pdb & 32.58 & d & 61\\
11 & 1r69.pdb & 11.40 & a & 63 & 1tig.pdb & 23.71 & d & 88\\
12 & 1ten.pdb & 10.74 & b & 90 & 1mjc.pdb & 21.99 & b & 69\\
13 & 1cyo.pdb & 10.55 & d & 88 & 1abo.a.pdb & 20.86 & b & 58\\
14 & 451c.pdb & 10.37 & a & 82 & 1fle.i.pdb & 19.94 & g & 47\\
15 & 1isu.a.pdb & 10.30 & g & 62 & 1hoe.pdb & 18.61 & b & 74\\
16 & 2ovo.pdb & 10.24 & g & 56 & 2wrp.r.pdb & 16.68 & a & 104\\
17 & 2bop.a.pdb & 9.97 & d & 85 & 451c.pdb & 16.29 & a & 82\\
18 & 1hoe.pdb & 9.36 & b & 74 & 1bhp.pdb & 16.08 & g & 45\\
19 & 1gvp.pdb & 9.31 & b & 87 & 1cmb.a.pdb & 15.63 & a & 104\\
20 & 1ptf.pdb & 9.19 & d & 87 & 1fxd.pdb & 14.79 & d & 58\\ \hline

\multicolumn{9}{l}{a: All alpha proteins}\\
\multicolumn{9}{l}{b: All beta proteins}\\
\multicolumn{9}{l}{d: Alpha and beta proteins (a+b)}\\
\multicolumn{9}{l}{g: Small proteins}\\
\hline
\end{tabular}
\caption{\sf The 20 proteins with highest
$\alpha$ value from both kernel folding potential and kernel design potential.
The residue number, and SCOP class also were listed.
}
\label{tab:highest_alpha}
\end{center}
\end{small}
\end{table}

\begin{figure}[tbh]
\centerline{\epsfig{figure=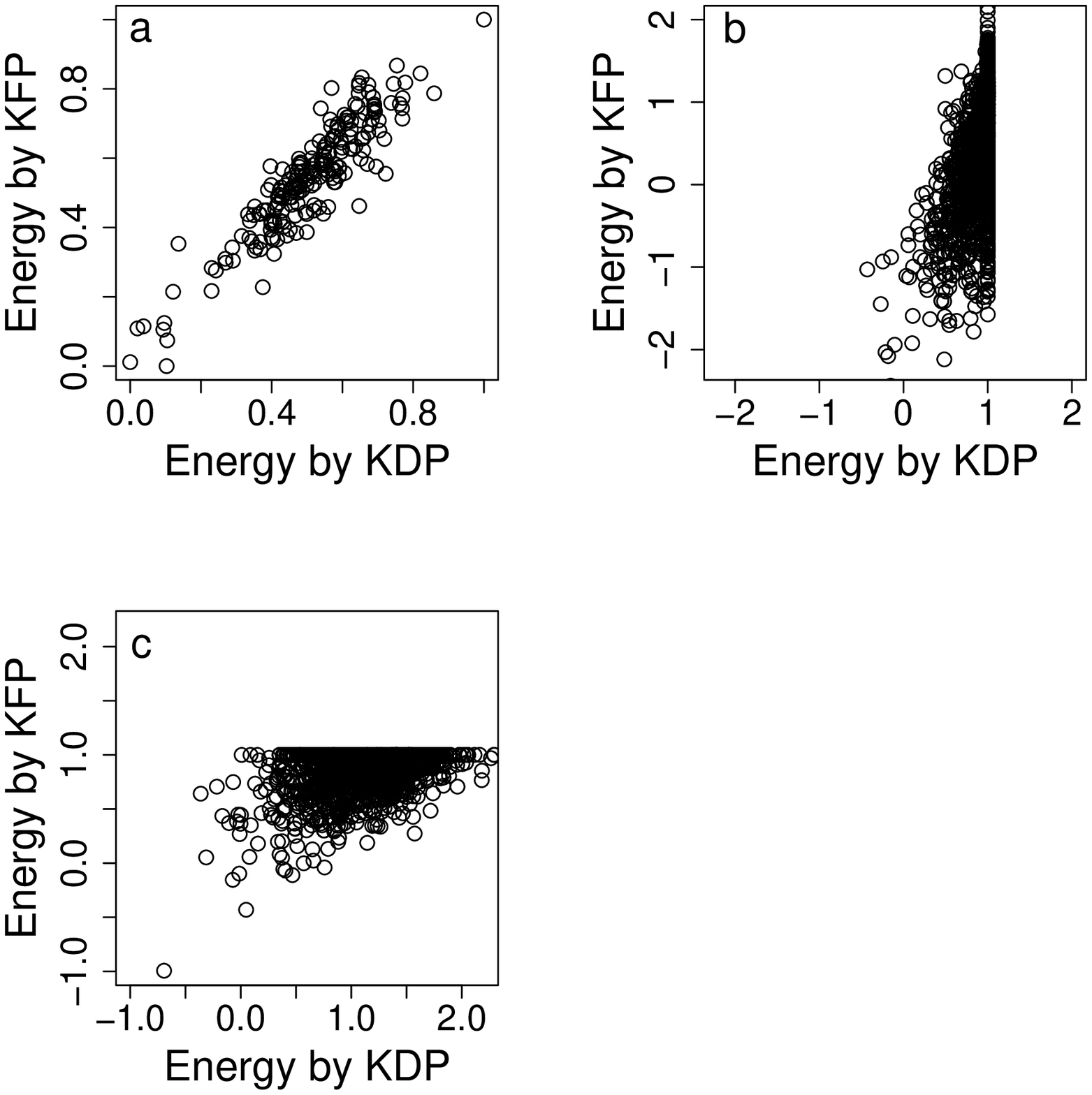,width=4in}}
\caption{
Comparison of nonlinear kernel
folding potential and kernel design potential generated by optimized
discrimination against decoys from gapless threading.  (a). The energy
values of the nonlinear folding and design potentials for the 210 unit
vectors are strongly correlated ($R = 0.91$).  (b). The energy values
by both design potentials and by folding potential for decoys that
enter the nonlinear design functions are poorly correlated. (c). The
energy values for decoys that enter the nonlinear folding functions
are also poorly correlated.
}
\label{Fig:unitVector}
\end{figure}

We also compare the energy value of the potential functions for each
of the 210 unit vector $\bc_1 = \{1, 0, \ldots, 0\}^T, \cdots$, and $
\bc_{210} = \{0, \ldots, 1\}^T$. We normalize these values so $\max
H(\bc_i) = 1$ for both potentials (Fig~\ref{Fig:unitVector}a).  There
is strong correlation ($R = 0.91$) for folding and design potentials.

However, other methods reveal that kernel folding and design
potentials are different.  One method is to compare the energy values
of a subset of decoy structures that are challenging. That is, we
compare energies of decoys with $\alpha_i \ne 0$.
Fig~\ref{Fig:unitVector}b shows that for decoys appearing in the
design potentials, there is little correlation in energy values
calculated by design potential and by folding potential.  Similarly,
there is no correlation between energy values calculated by folding
potential and by design potentials for the set of structure decoys
entering the design potential function (Fig~\ref{Fig:unitVector}c).
It seems that although the values of $\alpha_N$s are similar for the
majority of the native proteins, design potential and folding
potential can give very different energy values for some
conformations.  This suggests that the overall fitness for design and
folding potential may be somewhat different.

\paragraph{Comparison with Other Potentials.}
Potential functions derived from kernel models have the form of
$H(\bc) = \sum_{d \in \cal D}\alpha_d \cdot K(\bc, \bc_d) - \sum_{N
\in \cal N}\alpha_n \cdot K(\bc, \bc_n) $, which cannot be directly
compared with other potential functions of the form $H(\bc) = \bw
\cdot \bc$.  Here we follow the same approach as above and compare the
energy values of the unit vectors, the native proteins, and the decoys
using different potential functions.  The energy values by nonlinear
folding and design potentials have very little correlation with energy
values evaluated using either optimal linear folding potential
\cite{TobiElber00_Proteins_1} or statistical potential
\cite{Miyazawa96_JMB} (Fig~\ref{Fig:testSetenergies}c-f).  It seems nonlinear
potentials developed in this work contains information absent in other
potential functions.

\section{Discussion}

A basic requirement for computational studies of protein folding and
protein design is an effective potential function, which allows the
search and the identification of native structures and native
sequences.  Current empirical potential functions are based on
weighted linear sum of pairwise interactions.  A major flaw of
such potentials is that they cannot recognize the native structures of
a large number of proteins from alternative decoy conformations
\cite{Vendruscolo00_Proteins,TobiElber00_Proteins_1}.

There are several routes towards improving empirical potential
functions.  One approach is to introduce higher order interactions,
where three-body or four-body interactions are explicitly incorporated
in the potential function
\cite{Zheng97,Munson97,Betancourt99_PS,Rossi02_BJ}.  The effectiveness
of this approach is likely to be assessed in the future with large
scale validation studies similar to those carried out for pairwise
potentials
\cite{Vendruscolo00_Proteins,TobiElber00_Proteins_1,LuSkolnick01_Proteins}.
A different approach to improve empirical potential function is to
introduce nonlinear terms.  Recently, Fain {\it et al} uses sums of
Chebyshev polynomials upto order 6 for hydrophobic burial and each
type of pairwise interactions \cite{Fain02_PS}.

In this work, we propose a different framework for developing
empirical nonlinear protein potential functions.  Instead of using
nonlinear function for each term of pairwise interactions, we use a
set of simple Gaussian kernel functions located at both native
proteins and decoys as the basis set.  We regard the decoys as
equivalent to the reference state or null model used in statistical
potential. The expansion coefficients $\{ \alpha_N \}, N \in {\cal N}$
and $\{ \alpha_D \}, D \in {\cal D}$ of these Gaussian kernels
determines the protein function.  As long as the native proteins and
decoys are represented as unique vectors $\bc \in \real^d$, the Gram
matrix of the kernel function is full-rank.  Therefore, the kernel
function effectively maps the protein space into a high dimensional
space in which effective discrimination with a hyperplane is easier to
obtain.  The optimization criterion here is not $Z$-score, rather we
search for the hyperplane in the transformed high dimensional space
with maximal separation distance between the native protein vectors
and the decoy vectors.  This choice of optimality criterion is firmed
rooted in a large body of studies in statistical learning theory,
where expected number of errors in classification of unseen future
test data is minimized probabilistically by balancing the minimization
of the training error (or {\it empirical risk}) and the control of the
capacity of specific types of functions of potential function
\cite{Vapnik95,Burges98,ScholkopfSmola02}.

This approach is general and flexible, and can accommodate other
protein representation, as long as the final descriptor of protein
and decoy is a $d$-dimensional vector $\bc$.  In addition, different
forms of nonlinear functions can be designed using different kernel
functions, such as polynomial kernel and sigmoidal kernels.  It is
also possible to adopt different optimality criterion, for example, by
minimizing the margin distance expressed in 1-norm instead of the
standard 2-norm Euclidean distance.

A useful observation obtained from this study is that sequence decoys
obtained from gapless threading are quite challenging.  In fact, we
found that no linear potential function exists that can discriminate a
training set of 440 native sequence from sequence decoys generated by
gapless threading.  The success of nonlinear potential in perfect
discrimination of this training set native sequences and its good
performance in identifying the native sequences in an unrelated test
set of 194 proteins indicate that nonlinear kernel potential is a
general strategy for developing effective potential function for
protein sequence design.

It is informative to examine the three misclassified proteins by the
kernel folding potential ({\tt 1bx7}, {\tt 1hta}, and {\tt 3erd}).
Hirustasin {\tt 1bx7} contain five disulfide bonds, which are not
modeled explicitly by the protein description.  {\tt 1hta} (histone
Hmfa) exists as a tetramer in complex with DNA under the physiological
condition. Its nature structure may not be the same as that of a lone
chain. The two terminals of this protein are rather flexible, and
their conformations are not easy to determine. {\tt 3erd.a} (estrogen
receptor $\alpha$ ligand-binding domain) has extensive contacts with
ligand.  These unmodeled interactions may alter protein conformation.
Among the 14 native sequences misclassified by the kernel design
potential ({\tt 1a73.a}, {\tt 1bd8}, {\tt 1bea}, {\tt 1bm8}, {\tt
1bn8.a}, {\tt 1bvy.f}, {\tt 1cku.a}, {\tt 1dpt.a}, {\tt 1hta}, {\tt
1mro.c}, {\tt 1ops}, {\tt 1qav.a}, {\tt 1ubp.b}, {\tt 3ezm.a}),
several have extensive interchain interactions, although the contact
ratio is below the rather arbitrary threshold of 30\%: {\tt 1a73.a}
has Contact Ratio of $23\%$, {\tt 1mor.c} has $24\%$, {\tt 1upb.b} has
$19\%$, and {\tt 1qav.a} has $13\%$.  It is possible that the
substantial contacts with other chains would alter the confirmation of
the protein.  Amylase inhibitor {\tt 1bea} contains 5 disulfide bond
not explicitly modeled. {\tt 1cku.a} (electron transfer protein)
contains an iron/sulfur cluster, which covalently bind to four Cys
residues and prevent them from forming 2 disulfide bonds. These
covalnt bonds are not reflected in the protein description.  {\tt
1bvf} (oxidoreductase) is complexed with a heme and an FMN group. The
conformations of {\tt 1cku.a} and {\tt 1bvf.f} may be different upon
removing of these functionally important hetero groups. Altogether,
there are some rationalization for 10 of the 16 misclassified
proteins.

Our goal in this study is to explore an alternative formulation of
potential function and assess the effectiveness of this new approach
with experimental data. The nonlinear potential functions obtained in
this study should be further improved.  For example, unlike the study
of optimal linear potential \cite{TobiElber00_Proteins_1}, where
explicitly generated three-dimensional decoys structures are used in
training, we used only structure decoys generated by threading.  The
test results using the {\sc 4state\_reduced} set and the {\sc
lattice\_ssfit} are comparable or better with other residue-based
potential (see Fig~\ref{Fig:4state} and Table~\ref{tab:4state}).  It
is likely that further incorporation of explicit three-dimensional
decoy structures in the training set would improve the protein
potential.

In summary, we show in this study an alternative formulation of
protein function using a mixture of Gaussian kernels.  We demonstrate
that this formulation can lead to effective folding potential and
design potential that perform well in independent tests.  For protein
sequence design where challenging decoys are available from gapless
threading, nonlinear kernel potential can have perfect classification
in the training set of 440 proteins, while linear potential and
statistical potential failed. It also performs better in an
independent test set of 194 proteins and reduce the misclassification
to 40\% of that of optimal linear potential.  Our results suggest that
more sophisticated functional form other than the simple weighted sum
of contact pairs can be useful for studying protein folding and
protein design.  This approach can be generalized for any other
protein representation, {\it e.g.}, with descriptors for explicit
hydrogen bond and higher order interactions.

\section{Acknowledgment}
We thank Dr.\ Hao Li for very helpful comments on protein design, Dr.\
Yang Dai for discussion on linear programming, and Dr.\ M\'esz\'aros
for help in using the BPMD package.  Dr.\ Michele Vendruscolo kindly
provided the list of the set of 456 proteins.  This work is supported
by funding from National Science Foundation CAREER DBI0133856,
DBI0078270, and American Chemical Society/Petroleum Research Fund
35616-G7.

\section{Appendix}

\begin{lemma} 
For a potential function in the form of weighted
linear sum of interactions, a decoy always has energy values
higher than the native structure by at least an amount of $b>0$, {\it i.e.},
\begin{equation}
\bw \cdot (\bc_D - \bc_N) > b \quad \mbox{ for all }
\{(c_D - c_N)| D \in {\cal D} \mbox{ and } N \in {\cal N}\}
\label{proof1}
\end{equation}
if and only if the origin $\bf 0$ is not contained within
the convex hull of the set of points $\{ (\bc_D - \bc_N)
| D \in {\cal D} \mbox{ and } N \in {\cal N} \}$.
\end{lemma}

\begin{proof}
Suppose that the origin $\bf 0$ is contained within the convex hull
${\cal A} = \mbox{conv} (\{ \bc_D - \bc_N \})$ of $\{ \bc_D - \bc_N
\}$ and Equation (\ref{proof1}) holds.  By the definition of
convexity, any point inside or on the convex hull $\cal A$ can be
expressed as convex combination of points on the convex
hull. Specifically, we have:
\[
{\bf 0} = \sum_{(c_D- c_N) \in \cal A} \lambda_{c_D - c_N } \cdot(\bc_D - \bc_N), \quad
\mbox{ and } \sum \lambda_{c_D - c_N } = 1, \lambda_{c_D - c_N} >0.
\]
That is, we have the following contradiction:
\[
0 = \bw \cdot {\bf 0} = \bw \cdot \sum_{c_D- c_N}
 \lambda_{c_D - c_N }
 \cdot(\bc_D - \bc_N) 
  = \sum_{c_D - c_N}
 \lambda_{(c_D, c_N )} 
\cdot \bw \cdot (\bc_D - \bc_w)
> \sum_{c_D- c_N}
 \lambda_{c_D- c_N} 
\cdot b = b.
\]

Because the convex hull can be defined as the intersection of half
hyperplanes derived from the inequalities, if a half hyperplane has a
distance $b>0$ to the origin, all points contained within the convex
hull will be on the other side of the hyperplane \cite{Edels87}.
Therefore, $\bw \cdot (\bc_D - \bc_N) > b$ will hold for all
$\{ (\bc_D - \bc_N) \}$.
\end{proof}

\newpage

\bibliography{nigms,svm,array,prop,pack,prf,lattice,bioshape,liang,potential,career,pd02,pair,design}
\bibliographystyle{protein}

\end{document}